\documentstyle{article}

\newcommand\bg{\begin{eqnarray}}
\newcommand\ed{\end{eqnarray}}
\newcommand\bgn{\begin{eqnarray*}}
\newcommand\edn{\end{eqnarray*}}

\def\D{\partial}

\def\ra{\rightarrow}

\def\muhat{\hat{\mu}}
\def\inf{\infty}

\def \coma{~~~,}
\def \stop{~~~.}

\input epsf

\begin{document}
\begin{titlepage}

\title{\bf{
A New Approach to Numerical \\ Quantum Field Theory}}
\author{
Santiago Garc\' \i a ${\dag}$ {\thanks{Research supported in part by
 NSF Grant ASC-9211072}}
\\
Department of Physics \\
Brown University \\
Providence RI 02912 \\\\
G.S.~Guralnik
{\thanks{Research supported in part by DOE Grant DE-FG02-91ER40688 - Task D}}
{\thanks{Permanent Address:~Department of Physics, Brown University,
Providence RI 02912}}
\\
C-3 Los Alamos National Laboratory \\
Los Alamos NM, 87545 \\\\
John Lawson ${\dag}$ \\
Department of Physics \\
Brown University \\
Providence RI 02912}
\date{}
\maketitle
\begin{abstract}
\noindent
In this note we present a new numerical method for solving Lattice
Quantum Field Theory. This {\em{Source Galerkin Method}} is fundamentally
different in concept and application from Monte Carlo based methods
which have been the primary mode of numerical solution in
Quantum Field Theory.
Source Galerkin is not probabilistic and
treats fermions and bosons in an equivalent manner.
\end{abstract}
\thispagestyle{empty}

\vskip-20cm
\noindent
\phantom{bla}
\hfill{{\bf{BROWN-HET-908}}} \\
\end{titlepage}

\newpage
\section{Introduction}
\smallskip
\par
Over a decade ago, it was realized that
the algorithms and computer power existed to
calculate nonperturbative consequences of Quantum Field Theory
\cite{creutz}. Initial results using Monte Carlo
methods and the quenched approximation were a striking
confirmation of the power of numerical methods and the general
correctness of the ideas of lattice QCD \cite{monte0}. After countless
hours of calculation with ever more powerful computers
and vast improvements of the
initial class of algorithms, there is still the hope of
great success as machines become much faster. While there
has been demonstrable progress, it can be argued that important
but mostly incremental improvements in results have been made.

It is the purpose of this note to introduce a very different
class of approaches to numerical quantum field theory.
The new  methods, while
not without their own difficulties, have the promise of allowing
a great increase in the accuracy and speed of numerical
calculation. Particularly important  is that
fermions can be treated in essentially the same manner as
bosons.

For clarity, we will illustrate these ideas with very simple
examples.
Our initial
analysis was produced through the use of
ALJABR*  (a product of Fort Pond Research) which
is a very
reliable computer algebra program. It is not necessary to
resort to Fortran or C coded implementations of our methods
until quite complicated systems are studied. A good
workstation will produce a remarkable number of results.

Our approach is based on considering field theory in
the presence of external sources.
We start with the functional
equations for the vacuum amplitude $Z$.  On a lattice, these are
converted to a set of coupled linear differential
equations in the discretized sources $J$.

The high degree of
symmetry due to the translation and reflection invariance of the
lattice can be used to our advantage
in constructing the solution. For this note, $Z$ will be
approximated by a power series in $J$~\cite{inconsistency}.
Faster
convergence and the full power of our ideas is obtained
with other more structured approximations that will be presented
elsewhere.

Truncated approximations in multiple dimensions are intrinsically
inconsistent. One
particularly elegant method to handle this problem
is to require that weighted averages of the equation residual
vanish.  A clever choice of weight functions
speeds the convergence to the exact solution.
Techniques of this sort are familiar in the  numerical solution
of differential equations and are called Galerkin methods~\cite{fletcher}.
The
accuracy of the approximation can be
judged by measuring its numerical stability and convergence.

     Since our expansion and weight functions depend on the external
sources, we call this method {\em{the Source Galerkin method}}.
Except for
the anti-commutation of fermionic
sources, fermions and bosons can be treated symmetrically.
Nested solutions for dynamical
fermion problems become possible with existing computers.

\section{Functional Formulation in the Presence of Sources}

We begin by studying a scalar field
$\Phi(x)$ in the
presence of an external source $J(x)$.
As an example, consider a quartically self interacting scalar field
$\Phi$
on a periodic D-dimensional
lattice, with Euclidean action
\bg
S_{E}(\Phi,J)=
\sum_x \, {1 \over 2} \sum_y
\left( \Phi_x \left(-\Box + M^2 \right)_{x,y} \Phi_y \right) +
{g \over 4}\, \Phi_x^4 - J_x \Phi_x  \coma
\ed
where
\bg
\left(-\Box + M^2 \right)_{x,y} \equiv
\left(2 D + M^2 \right) \,\delta_{x,y} -
\sum_{\muhat}{\delta_{y,x+\muhat}} \stop
\ed
The lattice
spacing $a$ has been absorbed
in the definition of the sources and parameters, and
the $\muhat$ summation
is over nearest neighbors.
The generating function $Z(J)$ satisfies the
the following  set of coupled partial differential equations in
source space
\bg
\left( -\Box + M^2 \right)_{x,y}\, {\D Z \over \D {J_{y}}} +
\,g \, {{\D^3 Z  }\over {\D {J_{x}^3}}}\,
- J_{x} Z = 0 \stop
\label{eqlat}
\ed
There is one equation for each $J_{x}$.
The Euclidean (unconnected) lattice
Green's functions can be extracted from $Z(J)$ by differentiation
\bg
\langle \Phi(x_{1}), \ldots \Phi(x_{m}) \rangle =
G_{m}( x_{1},\ldots x_{m})\,=\,
{\D^{m} Z \over \D J_{x_{1}}  \ldots \D J_{x_{m}}}
 \Big\vert_{J=0}
\qquad m=1,2,3\ldots  \stop
\ed
For the rest of this note
we will examine only $D=1$ theories;
the ideas generalize to higher dimensions.
If fermions were present, the Grassman
nature of the equations
causes only technical complications.

\section{Boundary Conditions}
A straightforward way to attack (\ref{eqlat})
is to construct a power series solution~\cite{inconsistency}
\bg
Z(J_{1},J_{2},\ldots J_{N})=
\sum_{n_{1},n_{2},\ldots ,n_{N}} \,
G_{n_{1},n_{2},\ldots n_{N}} \,
J_{1}^{n_{1}} J_{2}^{n_{2}}\ldots J_{N}^{n_{N}}  \coma
\label{powser}
\ed
with $n_{i}, i=1,2,\ldots ,N$ nonnegative integers.

Before solving (\ref{eqlat}) for the coefficients
$ G_{n_{1},n_{2},\ldots n_{N}} $, we must specify boundary
conditions. Since~(\ref{eqlat}) is homogeneous in $Z$,
we can normalize the vacuum amplitude
\bg
Z(J=0)=1 \coma
\label{bound1}
\ed
and require odd Green's functions to vanish
\bg
\langle \Phi_{x_1 \ldots \Phi_{x_{n}}} \rangle =
{\D^n Z \over \D J_{x_{1}}
\ldots \D J_{x_{n}}}
\Big\vert_{J=0}  = 0 \qquad \qquad
\qquad \forall J_{x} {,}~n=1,3,5 \ldots  \stop
\ed
If equation (\ref{eqlat}) is considered as an initial value problem,
we need also to specify the second derivatives at the origin
\bg
\langle \Phi_{x} \Phi_{y} \rangle =
{\D Z \over \D J_{x}\D J_{y}} \Big\vert_{J=0}   \stop
\ed
This  appears to require a prior knowledge of the
two point Green's function \cite{cooper}.

Alternatively, we demand
the constructed $Z$ to approach smoothly the free field
generating functional
\bg
Z_{0}(J)=\displaystyle{e^{J_{x} (-\Box + M^2)^{-1}_{x,y} J_{y}}}   \coma
\ed
when $g \ra 0^{+}$.
This requirement is
an implicit boundary condition of
all Monte-Carlo (path integral) simulations and,
together with the above
conditions, uniquely determines $Z$.
Other classes of
solutions to the field equations (\cite{cooper}, \cite{Caianello},
\cite{gerrygarcia}) may also be of physical interest.

We found that one
way of implementing this boundary condition is
to truncate  $Z$  at some finite order  $K$
in the sources $J$, so that  all
unconnected Green's
functions with more than $K$ legs vanish.  Truncation then
introduces an
approximation scheme which can be systematically improved
by truncating at successively higher orders and
taking the limit $K \rightarrow \inf$. The convergence of this process
will be discussed in future publications.

\section{Lattice Symmetries}

The number of
coefficients in (\ref{powser})
grows exponentially with the number of sites and with
increasing truncation order \cite{counting}. Consequently,
this particular
version of the Source Galerkin method is only of
use for small lattices.
We can get somewhat further by exploiting symmetries
to reduce the number of unknowns to solve.  For instance,
in a lattice of ten sites, one can invoke translation
and reflection symmetries
and  group the 45 second order monomials of the form
$J_{x} J_{y}$ into five invariant classes,
thus reducing considerably the number of unknowns
(see (\ref{invpol})).

Since the differential operators in (\ref{eqlat}) are
closed under the action of the lattice symmetry group,
it is natural to look only for invariant solutions~$Z$.
\bg
Z (J) = \sum_{\sigma} \sum_{n}
{a}_{\sigma}^{(n)}{P}_{\sigma}^{(n)} (J)  \coma
\label{expan1}
\ed
where ${P}_{\sigma}^{(n)}(J)$
are invariant polynomials of order $n$
corresponding to a particular symmetry class $\sigma$.
The number of
invariant classes for a given order will depend on both the
number of lattice sites and on the number of symmetry operations
for a given lattice.  Details of how to construct
these invariant polynomials will be given elsewhere, but it is
important to realize now that they form a {\em{complete}} set for the
class of lattice invariant solutions we are considering.
We believe the ability to easily exploit symmetries is a major advantage
of our approach.

    Once  Z  is constructed as lattice symmetric, it is sufficient to
use only one equation from the coupled set (\ref{eqlat}):  all
equations in the coupled system express identical dynamics and only
differ by lattice translations and reflections.

\section{The Galerkin method}

When solving a differential equation by power series
methods,
we substitute the series into the equation.  Equating
like terms in the resulting polynomial yields a system of linear
algebraic equations that determine the expansion
coefficients of
the original power series.

Applying this procedure to (\ref{eqlat}) and the power series
(\ref{expan1}) for $Z$ yields
an inconsistent set of linear equations
for the truncated Green's functions:
there are more independent
equations than unknowns ${a}_{\sigma}^{(n)}$.
The Galerkin method is particularly suited to deal with this problem.

The key features of the Galerkin method can be stated concisely
\cite{fletcher}. Assume that
we have a problem governed by a linear differential equation
\bg
L\,f = b \stop
\label{eqitem}
\ed
$L$ is a linear differential
operator,
and $b$ a known function of the $N$ variables
${\bf{x}} \equiv (x_{1}{,}\ldots ,x_{N}) $.
The solution is sought
in a N-dimensional domain ${\cal{D}}$.
The Galerkin method assumes that the exact solution
$f=f({\bf{x}})$ is
accurately represented by an approximate $f^{*}_{K}$
\bg
f^{*}_{K}({\bf{a}};{\bf{x}})=\,\varphi_{0}({\bf{x}})\,
+\sum_{j=1}^{K}{a_{j} \,\varphi_{j}} ({\bf{x}})  \coma
\label{trialgal}
\ed
where $\varphi_{0}$ is introduced to satisfy the boundary conditions and the
$a_{j}$'s are coefficients to be determined. The $K$ functions
$\varphi$ are selected according to some criterion,
like knowledge of the approximate behavior of the solution or mathematical
completeness. These are called {\it{basis}}, or {\it{trial}}, functions.
Substitution of (\ref{trialgal})
in (\ref{eqitem}) produces a nonzero residual $R$ given by
\bg
R({\bf{a}};{\bf{x}})=\,L(f^{*})\,=\,
L(\varphi_{0})
+\sum_{j=1}^{K}{a_{j} \,L(\varphi_{j})}  \stop
\label{residual}
\ed
The unknown coefficients $a_{j}$ are determined
by solving the following system of {\em{linear}} algebraic equations
\bg
< \bar{\varphi_{k}}\, , R > =
< \bar{\varphi_{k}}\, , b >
\coma ~~~~~~~~~~~~~~~~~~~~~~~~~~~~~~~~~~~~~~
\\ \nonumber
\longrightarrow \quad
 \sum_{j=1}^{K} < \bar{\varphi_{k}}, L \,\varphi_{j} >{a_{j}}
= < \bar{\varphi_{k}}, b > -< \bar{\varphi_{k}}, L \,\varphi_{0} > \coma
 \qquad k=1,2,\ldots K  \coma
\label{test}
\ed
where $\bar{\varphi}$ are suitable {\em{test}} functions,
and $<\,{,}\,>$ is the inner (scalar) product
\bg
<\,\phi{,}\,\chi\,>\,\equiv\,
\int_{{\cal{D}}}{\,d^{N} x \,\phi\,\chi} \stop
\ed
In the pure Galerkin method,
the test functions
are chosen from the same functional space
as the basis functions $\varphi \,$ in (\ref{trialgal}).

Under very general conditions \cite{fletcher} it is possible to prove
that the Galerkin solution $f_{K}^{*}$ converges to the true solution
$f$ {\em{in the mean}}
\bg
\lim_{K \rightarrow \inf} \int_{{\cal{D}}}
\vert \vert f_{K}^{*} -f \vert \vert \,= 0  \stop
\ed

\section{Example}
As an illustration of the Galerkin method
with truncated boundary conditions,
we solve (\ref{eqlat}) for the case of a one dimensional lattice
with three sites ($D=1$ and $N=3$ in (\ref{eqlat})),
expanding the generating function $Z$ up to the first
non-trivial order in the sources, that is, up to quartic
invariant polynomials.

First, we construct $Z_{4}^{*}=Z_{4}^{*}
(J_{1},J_{2},J_{3})$ as a linear combination
of lattice invariant polynomials.
\bg
Z_{4}^{*}(J_{1},J_{2},J_{3})= 1 + a_{0}^{(2)}  P_{0}^{(2)}(J)+
a_{1}^{(2)} P_{1}^{(2)}(J)
+ a_{0}^{(4)} P_{0}^{(4)}(J) \ldots
+ a_{3}^{(4)} P_{3}^{(4)}(J)  \coma
\ed
with
\bg
P_{0}^{(2)}(J)=J_{1}^2+J_{2}^2+J_{3}^2  \coma
\\\nonumber
P_{1}^{(2)}(J)= J_{1} J_{2}+J_{2} J_{3}+J_{3} J_{1}  \coma
\\\nonumber
P_{0}^{(4)}(J)= J_{1}^{4} +J_{2}^{4} +J_{3}^{4}  \coma
\\\nonumber
\ldots \stop ~~~~~~~~~~~~
\label{invpol}
\ed
Only six $P_{\sigma}^{(n)}$
are needed.
The boundary conditions in section $3$
have already been imposed.
We choose as our integration domain ${\cal{D}}= [-\epsilon, \epsilon]^{3}$, a
hypercube of size $2 \epsilon$ centered at the origin, with $\epsilon$
infinitesimal and set to zero {\em{after}} the calculation is done (this
is of course possible only with algebraic languages like ALJABR, otherwise
a numerical limit has to be taken).

Because the operator in (\ref{eqlat}) is odd and ${\cal{D}}$ is
symmetric around $J=0$, we need odd
test functions in order to obtain non-vanishing inner products. A
prescription
for test functions that we found numerically convenient, since
it provides the necessary number of independent equations for
the coefficients $a_{\sigma}^{(n)}$, is
the derivative of the basis polynomials respect to a
source $J_{x}$
\bg
T_{\sigma}^{(n)} = {\D P_{\sigma}^{(n)} \over \D J_{x}} \stop
\label{testf}
\ed
Lattice symmetry makes any choice of $x$ equivalent.
We set $x=1$. Any other set
of test functions that produces the correct number of independent equations
will also work, but with a different rate of convergence.

The Galerkin method is then applied to the residual of (\ref{eqlat}).
The  resulting
coefficients $a_{\sigma}^{(n)}$ are rational
functions of the mass and the coupling. For instance, the
Galerkin approximation to
$\langle \Phi(0) \Phi(0) \rangle $ is
\bg
{\langle \Phi(0) \Phi(0) \rangle}^{*} =
{\D Z_{4}^{*} \over \D J_{1}^{2}} \Big\vert_{J=0}  =
2 a_{0}^{(2)} \equiv {A(M,g) \over B(M,g)} \coma
\label{green2}
\ed
with
\bg
A(M,g)=
M^2 (M^2+1) (M^2+3) (266 M^4+929 M^2+788) + \\\nonumber
35 g (M^2+1) (M^2+2) (14 M^2+27) \coma ~~~~~~~~~~~
\ed
\bg
B(M,g)=  M^4 (M^2+3)^2 (266 M^4+929 M^2+788)+~~~\\\nonumber
 g (1288 M^8+7778 M^6+17997 M^4+18759 M^2+7092)+\\\nonumber
105 g^2 (M^2+1) (14 M^2+27) \coma ~~~~~~~~~~~~~
\ed
One can check that the above expression (\ref{green2})
has the right free field limit
when $g \rightarrow 0^{+}$.

The convergence of this process is illustrated in Table $1$,
where we show our results for
the two point function
$$G_{i,j}= \langle \Phi_i \Phi_j \rangle =
{{{\partial}^2 Z}\over { {\partial J_i}\partial J_j}}
\Big\vert_{J=0} \coma  $$
on an $11$ site lattice
with $M=1$ and $g=0.5$. We see that the results agree extraordinarily
well with the Monte-Carlo simulation,
and that the convergence is oscillatory \cite{gerrygarcia}.

\section{Conclusions}
We have presented a new method to numerically
attack problems in lattice QFT.  It is not probabilistic
and possesses extreme flexibility,
allowing for maximum input of information
(symmetry, spectral properties, etc.). Systematic
improvements of the approximations are possible, forcing
convergence (in the mean)  to the exact solution. We have illustrated
the method by looking at truncated Taylor series expansions.
These lose tractability for big systems. We will present more useful
nested approximations elsewhere.

     The differential formulation requires
treatment of boundary conditions. Truncation (among other boundary
conditions \cite{gerrygarcia}) guarantees
that the solution obtained from the differential equations
corresponds to the solution of the path integral. Other types
of solutions can also be analyzed with our method.

     Finally,  the Source Galerkin method gives real hope that
lattice fermion theories will yield to numerical treatment.  Monte
Carlo approaches in QCD involving fermions are problematic.
Calculation of the
fermion determinant makes inclusion of dynamical quarks in lattice
QCD prohibitive.  Simulations of strongly correlated systems in
condensed matter physics are afflicted by the minus sign problem
(\cite{johngerry}). We expect Source Galerkin
to be a powerful tool
for studying these systems since the above problems do not affect our
method.  The difference in computational complexity between bosons
and fermions is minimal. Indeed, part of the strength of our method
is the symmetry of approximations for bosons and fermions.
\leftline{}
\leftline{}
\leftline{}
\leftline{}
{\bf{Acknowledgements}}
G.S. Guralnik would like to thank Vance Faber for support and valuable
conversations during the course of this work. He is grateful to C division
for providing the environment that made the evolution of these ideas
possible. S. Garcia would like to thank Vance Faber for support during
a summer visit to Los Alamos. We have all benefited from conversations
with many of our collegues and particularly Greg Kilcup and
Stephen Hahn as we learned to
apply Galerkin ideas to quantum field theory. Jim O'Dell provided a site
license for the symbolic program ALJABR and considerable help in its
application to our ideas.

\newpage
\begin{table}[ht]
\centerline{Table 1}
  \begin{center}
   \begin{tabular}{||c|c|c|c|c|c||}
   \hline \hline
      \hphantom{xxx}       $i-j$     \hphantom{xxx}   &
      \hphantom{xxx}       $J^4$     \hphantom{xxx}   &
      \hphantom{xxxxx}     $J^6$     \hphantom{xxxxx} &
      \hphantom{xxxx}      $J^8$    \hphantom{xxxx}  &
      \hphantom{xxxx}      $Shanks$    \hphantom{xxxx}  &
      \hphantom{xxx}       $ M-C $   \hphantom{xxx}  \\ \hline
0 & 0.326538 &   0.360178 &    0.347223 & 0.35082   &   0.3510  $\pm$ 0.0005 \\
1 & 0.094508 &   0.117673 &    0.108121 & 0.11091   &   0.1110  $\pm$ 0.0004 \\
2 & 0.026768 &   0.038823 &    0.033422 & 0.03509   &   0.0351  $\pm$ 0.0003 \\
3 & 0.007454 &   0.012940 &    0.010269 & 0.01114   &   0.0112  $\pm$ 0.0004 \\
4 & 0.002075 &   0.004461 &    0.003197 & 0.00363   &   0.0036  $\pm$ 0.0003 \\
5 & 0.000694 &   0.001922 &    0.001218 & 0.00147   &   0.0011  $\pm$ 0.0004 \\
   \hline \hline
   \end{tabular}
   \end{center}
 \label{tablesh11}
\caption{Results for the two point function $G_{i,j}$
in a N=11 lattice (anharmonic
oscillator). Columns: $|i-j|$,
4th., 6th. and 8th. order Galerkin approximations,
numerical extrapolation (Shanks transformation)
and Monte-Carlo data with statistical error.}
\end{table}


\begin{thebibliography}{99}
\bibitem{inconsistency}{Attempts to find a Taylor series solution
around $J=0$
can fail \cite{lancosz} if the differential operator is singular (
coefficients of the
derivatives vanish or blow up at $J=0$) or if one introduces discontinuities
in the boundary conditions.
Since this is not the case
for the differential operator in (\ref{eqlat}), solutions
that admit a Taylor expansion in $J$ exist.}
\bibitem{multiple}{The issue of multiple solutions of the Schwinger-Dyson
equations has already been addressed in the literature, for instance
in \cite{Caianello} and \cite{cooper}.}
\end{thebibliography}
\end{document}